# Anomalous microwave conductivity coherence peak in c-axis oriented MgB$_2$ thin films

B.B. Jin[1], T. Dahm[2], A.I. Gubin[3], Eun-Mi Choi[4], Hyun Jung Kim[4], Sung-IK Lee [4], W.N.Kang [5], and N. Klein [1]

1) Forschungszentrum Jülich GmbH , Institute of Thin Films and Interfaces, D-52425 Jülich, Germany

2) Universität Tübingen, Institut für Theoretische Physik, Auf der Morgenstelle 14, 72076 Tübingen, Germany

3) Usikov Institute of Radiophysics and Electronics, NAS of Ukraine, 12 Acad. Proskura Str., 61085 Kharkov, Ukraine

4) National Creative Research Initiative Center for Superconductivity, Department of Physics, Pohang University of Science and Technology, Pohang, 790-784, Republic of Korea

5) Department of Physics, Pukyung National University, Pusan, 608-737, Korea

*Abstract*----The temperature dependence of the real part of the microwave complex conductivity at 17.9 GHz obtained from surface impedance measurements of two c-axis oriented MgB$_2$ thin films reveals a pronounced maximum at a temperature around 0.6 times the critical temperature. Calculations in the frame of a two-band model based on Bardeen-Cooper-Schrieffer (BCS) theory suggest that this maximum corresponds to an anomalous coherence peak resembling the two-gap nature of MgB$_2$. Our model assumes there is no interband impurity scattering and a weak interband pairing interaction, as suggested by bandstructure calculations. In addition, the observation of a coherence peak indicates that the π-band is in the dirty limit and dominates the total conductivity of our films.

Superconductivity in magnesium diboride ($MgB_2$) with a remarkable high critical temperature $T_c$ of 39K has attracted great attention both regarding its fundamental physics as well as practical applications[1-14]. Many theoretical calculations and experimental results have demonstrated its phonon-mediated pairing mechanism and s-wave pair symmetry. However, unlike conventional superconductors it displays a two energy gap structure. Combining a simple crystal structure and a high $T_c$ [15,16], it provides a real system to study the various fundamental properties related to two-band superconductivity [13,14].

In the present work we report on experimental results for the microwave conductivity of c-axis oriented thin films. For conventional superconductors this quantity displays a peak close to $T_c$ in its temperature dependence, which is called "coherence peak"[17][18]. The observation of its analog -the Hebel-Slichter peak- in the nuclear magnetic resonance (NMR) relaxation rate and its natural occurrence within the Bardeen-Cooper-Schrieffer (BCS) theory of superconductivity are regarded as one of the key triumphs of BCS theory [19]. Within BCS theory the peak arises due to the density of states (DOS) singularity which develops below $T_c$ when the superconducting gap opens. At lower temperatures an exponential decrease of the conductivity and the NMR relaxation rate sets in due to the condensation of quasiparticles. Usually the peak occurs at a normalized temperature around $t = T/T_c$ = 0.8-0.9 and can be observed both in the NMR relaxation rate and in the microwave conductivity, because both quantities possess the same coherence factors. However, the peak in the microwave conductivity is visible most clearly only in the dirty limit and gradually disappears when the mean free path becomes longer than the coherence length [20]. In the high-$T_c$ cuprates the behavior of these two quantities is quite different. In the NMR relaxation rate no Hebel-Slichter peak has been observed, while in microwave conductivity experiments a exceptionally high maximum occurs at temperatures much below $T_c$ [21][22] with its position being strongly dependent on frequency [23]. The absence of the Hebel-Slichter peak in NMR experiments was attributed to the d-wave nature of the superconducting order parameter in the cuprates [24-25]. The large peak in the conductivity has been interpreted as occurring due to an unusual rapid drop of the inelastic quasiparticle scattering rate below $T_c$ which one expects, if inelastic scattering is dominated by electron-electron scattering (e.g. spinfluctuations)[26]. Since the NMR relaxation rate is much less sensitive to the quasiparticle scattering rate, this scenario can explain the difference of NMR and microwave conductivity in the cuprates.

In this letter we report on the observation of an unusual conductivity maximum in microwave absorption measurements on $MgB_2$ and on a comparison with theoretical calculations based on a two-gap BCS-model. The temperature dependence of the effective surface impedance $Z_s^{eff}=R_s^{eff}+j\omega\mu\lambda^{eff}$ of two c-axis oriented $MgB_2$ films (samples S1210 and S1211) was measured at a frequency $\omega/2\pi$ of 17.9 GHz employing a sapphire resonator technique[7]. The films were fabricated using a two-step method by pulsed laser deposition. The detailed process is described in Ref.[27]. In our previous work the zero temperature gap ratio $\Delta(0)/kT_c$ and penetration depth $\lambda(0)$ were extracted using the clean limit BCS model with a $\Delta(0)/kT_c$ value being much lower than the BCS prediction of 1.76. Subsequently, the surface resistance $R_s$ and reactance $X_s$ can be calculated by impedance transformations[28]. Since the coherence length in our samples is much smaller than the penetration depth, the local limit applies and the real part of the complex microwave conductivity $\sigma_1$ can be derived from $R_s$ and $X_s$ using the formula $\sigma_1=2\omega\mu R_s X_s/(R_s^2+X_s^2)^2$. At $T_c$ we determined normal state conductivities of 0.137/$\mu\Omega$ cm (sample S1210) and 0.049/$\mu\Omega$ cm (sample S1211). We can get a rough estimate of the scattering rates $\Gamma$ in our films using the relation $\sigma=\varepsilon_0\omega_p^2\hbar/\Gamma$, where a plasma frequency of $\omega_p=5.9$ eV is used as suggested by bandstructure calculations[11]. According to this procedure we find $\Gamma\approx34$ meV (S1210) and $\Gamma\approx95$ meV (S1211). These values are bigger than the larger of the two gaps $\Delta_\sigma \approx 7$ meV, which suggests an analysis of our data within the dirty limit. In addition, many of the experiments and calculations mentioned above showed that MgB$_2$ appears to be a two gap superconductor. Therefore, in the present work we extracted $\Delta(0)/kT_c$ and $\lambda(0)$ values using the two gap model and the dirty limit. In this case the temperature dependence of the penetration depth is given by[29,30,31]:

$$\frac{\lambda^2(0)}{\lambda^2(T)} = \frac{\sigma_{n,\pi}\Delta_\pi(T)\tanh\frac{\Delta_\pi(T)}{2k_BT} + \sigma_{n,\sigma}\Delta_\sigma(T)\tanh\frac{\Delta_\sigma(T)}{2k_BT}}{\sigma_{n,\pi}\Delta_\pi(0) + \sigma_{n,\sigma}\Delta_\sigma(0)} \quad (1)$$

Here, $\sigma_{n,\sigma}$ and $\sigma_{n,\pi}$ are the partial normal state conductivities of the σ- and π-band, respectively. The temperature dependences of the two gaps $\Delta_\sigma(T)$ and $\Delta_\pi(T)$ are obtained from a solution of the two-by-two gap equation, as pointed out below.

Fig.1 shows the measured change of $\delta\lambda^{eff}$ (symbols) and fits based on Eq. 1 (lines). We found $\Delta_\pi(0)/kT_c = 0.69$ and $\lambda(0)=82$nm for sample S1210, and $\Delta_\pi(0)/kT_c = 0.79$ and $\lambda(0) = 118$nm for sample S1211. The values for the small gap are fairly consistent with observations from tunneling experiments [2-3]. Using these values, $\sigma_1$ was obtained by the procedure mentioned above. In Fig.2 we show the temperature dependence of $\sigma_1$ normalized to its value at $T_c$. A distinct feature of the temperature dependence of $\sigma_1$ is a maximum appearing around $t = 0.6$ rather than close to $T_c$, as one would expect from a BCS coherence peak.

In order to verify our measurement and extraction procedure, we also measured a 200nm-thick niobium film. The film was deposited on a silicon substrate at room temperature by DC magnetron sputtering in argon atmosphere. $T_c$ (=9.2K) and normal state resistivity $\rho$ (=4.4μΩ cm at 10K) were measured with standard four-probe method[32]. From the $T_c$ and $\rho$ values we can obtain the intrinsic coherence length $\xi_0=40$nm and mean free path l=10nm. This result shows that the dirty limit is a good approximation for our Nb thin film. Therefore, the same procedure was used to fit the measured change of penetration depth. Good agreement was obtained for $\Delta(0)/kT_c$=1.90 and $\lambda(0)$=91nm. Then, $\sigma_1$, as shown in Fig.1 (triangles), was extracted using the same procedures as above. A pronounced coherence peak appears close to T$_c$, as one expects for a conventional superconductor. This result is reproduced very well by a calculation based on the BCS model and Mattis-Bardeen conductivity with the parameters above, as shown in Fig.1 (solid line)[33]. Since the observation of the coherence peak in the microwave region needs highly accurate measurement of $R_s$ and $X_s$, results were reported only recently for the conventional superconductors Nb and Pb [17,18]. The success of our technique for the Nb thin film is a strict test because $\delta\lambda^{eff}$ and $R_s$ for MgB$_2$ are larger than for the Nb thin film.

Now, we turn back to the question about the nature of the maximum observed in our measurement on the MgB$_2$ samples. As discussed before, in the high-T$_c$ cuprates a much larger maximum below T$_c$ was attributed to a rapid drop of the scattering rate. Therefore one might be tempted to look for the same reason in this case. However, we do not expect a rapid drop of the scattering rate below T$_c$ in MgB$_2$. Measurements of the dc resistance can be well fitted by the Bloch-Grüneisen formula, which suggests that the normal-state transport properties are well described by an electron-phonon interaction [8]. At $T_c$ the dc resistance is already in the saturation regime, where impurity scattering dominates. Therefore, we cannot expect an important contribution of electron-electron scattering below $T_c$. Also, bandstructure calculations have shown that electron-phonon interaction and impurity scattering are sufficient to understand the properties of MgB$_2$ [10,11,14]. For these reasons the scenario of a rapid drop of the scattering rate in MgB$_2$ is rather unlikely as an explanation of the observed conductivity maximum.

If the maximum is related to coherence effects we should explain it does not appear close to $T_c$. For this reason we have calculated the temperature dependence of $\sigma_1$ within the two-band BCS model using the two-band gap equation [10] and the dirty limit Mattis-Bardeen formula for the conductivity[33]. In the two-band model the two superconducting gap values on the two Fermi surfaces (σ-band and π-band) have to be calculated from the following two by two matrix equation:

$$\Delta_\alpha = \sum_\beta \lambda^{\alpha\beta} \Delta_\beta \int_0^{\omega_c} dE \frac{\tanh \frac{\sqrt{E^2 + \Delta_\beta^2}}{2T}}{\sqrt{E^2 + \Delta_\beta^2}} \qquad (2)$$

where $\alpha$ and $\beta$ run over the two bands, $\omega_c$ is a characteristic phonon frequency, $\lambda^{\alpha\beta}=V^{\alpha\beta}N_\beta$ is a two-by-two coupling matrix, $N_\beta$ is the partial density of states of band $\beta$ and $V^{\alpha\beta}$ the symmetric matrix of the pairing interactions. From this equation we calculate the temperature dependence of the two gaps $\Delta_\sigma$ and $\Delta_\pi$ numerically using $\omega_c$ = 56 meV and $N_\pi/N_\sigma$=1.35 as given in Ref. [10]. In order to further reduce the number of free parameters, the pairing matrix is adjusted to give the experimental $T_c$ and the value of $\Delta_\pi(0)/kT_c$ of the small gap at zero temperature obtained from the fit to the penetration depth data in each sample. With these constraints given, there is only one free parameter left in Eq. (2): the interband pairing interaction $V^{\sigma\pi}$. In Fig. 3 we show the temperature dependence of the two gaps calculated this way using a ratio $r = V^{\sigma\pi}/V^{\sigma\sigma}$ of 0.25. The temperature dependence of the energy gaps resembles the experimental results by Gonnelli *et al.*[2]

Once the temperature dependences of the two gaps are known, the partial conductivities of the two bands can be calculated using the dirty limit Mattis-Bardeen formula [33]:

$$\frac{\sigma_\alpha(\omega)}{\sigma_{n,\alpha}} = \frac{1}{2\omega} \int_{-\infty}^{\infty} d\Omega \left( \tanh \frac{\Omega+\omega}{2T} - \tanh \frac{\Omega}{2T} \right) [N_\alpha(\Omega)N_\alpha(\Omega+\omega) + M_\alpha(\Omega)M_\alpha(\Omega+\omega)] \quad (3)$$

where

$$N_\alpha(\Omega) = \text{Re}\left\{\frac{|\Omega|}{\sqrt{\Omega^2 - \Delta_\alpha^2}}\right\} \quad \text{and} \quad M_\alpha(\Omega) = \text{Re}\left\{\frac{\Delta_\alpha \, \text{sgn}(\Omega)}{\sqrt{\Omega^2 - \Delta_\alpha^2}}\right\}$$

are the normal and anomalous densities of states for the two bands. In Ref. [11] it has been shown that interband impurity scattering is expected to be small in $MgB_2$, because the two bands possess different symmetries, which strongly reduces the scattering matrix element for interband impurity scattering. In this case the total conductivity $\sigma_1(\omega)$ is just the sum of the partial conductivities $\sigma_\pi(\omega)$ and $\sigma_\sigma(\omega)$. In order to calculate the temperature dependence of $\sigma_1$ for our measurement frequencies we need to know the ratio of the partial conductivities in the normal state. This ratio depends on the plasma frequencies $\omega_{p,\alpha}$ and the scattering rates $\Gamma_\alpha$ in the two bands $\sigma_{n,\pi}/\sigma_{n,\sigma} = \omega^2_{p,\pi}\Gamma_\sigma / \omega^2_{p,\sigma}\Gamma_\pi$. From the analysis in Ref.[11] we know that $\omega^2_{p,\pi}/\omega^2_{p,\sigma} \approx 2$, however, the ratio of the scattering rates will depend on the impurities and imperfections present in our film and remains unknown. We will show in the following that we can obtain a coherence peak appearing significantly below $T_c$, if we assume that the total conductivity is dominated by the partial conductivity of the π-band, i.e. $\sigma_{n,\pi} >> \sigma_{n,\sigma}$. In the other limiting case that the σ-band dominates the total conductivity we find a conventional coherence peak appearing close to $T_c$, which does not fit our observation.

Assuming $\sigma_1(\omega) = \sigma_\pi(\omega)$ we show the temperature dependence of $\sigma_1$ at 17.9 GHz obtained from Eqs. (2) (solid lines) for both samples. The best fit was found for $r = 0.25$ for both samples. This value is in fair agreement with the value of $r = 0.17$ from bandstructure calculations in Ref. [10].

We can obtain an intuitive understanding of the peak position, if we remember that the coherence peak appears when the condensation of quasiparticles sets in, which roughly appears when $\Delta(T) \approx k_B T$. In Fig. 3 this criterion is fulfilled when the dashed line intersects the small gap (arrow). From this picture it becomes clear that the downward shift of the coherence peak is directly related to the smallness of the gap.

To summarize, we have presented experimental results of the temperature dependence of the microwave conductivity in c-axis oriented $MgB_2$ thin films. We find an anomalous peak around $t = T/T_c = 0.6$ far below the temperature at which one would expect to see the BCS coherence peak. We argue that this peak is not related to a rapid drop in the quasiparticle scattering rate below $T_c$, as for the high-$T_c$ cuprates. Instead, we provide arguments that this anomalous peak in $MgB_2$ is related to the two-gap nature of its superconducting state, particularly to the smaller gap. Furthermore, our results imply that the π-band is in the dirty limit and dominates the total conductivity of our films.

One of the authors (A.I.Gubin) is funded in part within the INTAS programme by the European Union, which supports his research stay in Jülich. The work at Postech was supported by the Ministry of Science and Technology of Korea through the Creative Research Initiative Program.

[1]  J. Nagamatsu J, N.Nakagawa, T. Muranaka, Y.Zenitani and J.Akimitsu, *Nature* **410,** 63 (2001).
[2]  R. S. Gonnelli et.al., *Phys. Rev. Lett.* **89**, 247004 (2002).
[3]  H. Schmidt, J.F.Zasadzinski, K.E.Gray, and D.G.Link, *Phys. Rev. Lett* **88**, 127002 (2002).
[4]  S. Souma et.al., *Nature* **423**, 65(2003).
[5]  F. Bouquet, R.A. Fisher, N.E. Phillips, D.G.Hinks, and J.D.Jorgensen, *Phys. Rev. Lett*. **87**, 047001 (2001).
[6]  H.D.Yang, et.al., Phys. *Rev. Lett.* **87**, 167003 (2001).
[7]  B. B. Jin et. al., *Phys. Rev. B* **66**, 104521 (2002).
[8]  Kijoon H.P.Kim et al., *Phys. Rev. B* **65**, 100510(R) (2002).
[9]  J.Kortus, I.I.Mazin, K.D.Belashchenko, V.P.Antropov, and L.L.Boyer, *Phys. Rev. Lett.* **86**, 4656 (2001).
[10]  A.Y.Liu, I.I.Mazin, and J. Kortus, *Phys. Rev. Lett*. **87**, 087005 (2001).
[11]  I.I.Mazin, O.K.Andersen, O.Jepsen, O.V.Dolgov, J. Kortus, A.A.Golubov, A. B. Kuz´menko, and D. Van der Marel, *Phys. Rev. Lett.* **89**, 107002(2002).
[12]  A. I. Posazhennikova, T. Dahm and K. Maki, *Europhys. Lett.,* **60**, 134 (2002).
[13]  W.Pickett, *Nature*, vol.418, 733(2002).
[14]  H.J.Choi, D. Roundy, H.Sun, M.L.Cohen, and S.G.Louie, *Nature*, vol.418, 758(2002).
[15]  S.V. Shulga et.al., *Phys.Rev.Lett.* **80,** 1730(1998).
[16]  Hyeonjin Doh, M.Sigrist, B.K.Cho and Sung-IK Lee, *Phys. Rev. Lett.* **83**, 5350(1999).
[17]  O. Klein, E.J.Nicol, K.Holczer and G.Grüner, *Phys. Rev. B.* **50**, 6307 (1994).
[18]  M.R.Trunin, A.A.Zhukov, and A.T.Sokolov, *JETP* **84**, 383(1997).
[19]  M. Tinkham, "*Introduction to superconductivity*", 2$^{nd}$ ed. (McGraw-Hill, New York, 1996).
[20]  F. Marsiglio, *Phys. Rev. B* **44**, 5373 (1991).
[21]  D. A. Bonn, P. Dosanjh, R. Liang and W. N. Hardy, *Phys. Rev. Lett.* **68**, 2390 (1992).
[22]  M. C. Nuss, P. M. Makiewich, M. L. O´Malley, E. H. Westerwick, P. B. Littlewood, *Phys. Rev. Lett.* **66**, 3305 (1991).
[23]  A. Hosseini et.al., *Phys. Rev. B.* **60**, 1349 (1999).
[24]  P.C. Hammel et al., *Phys. Rev. Lett.* **63**, 1992 (1989).
[25]  N. Bulut and D. J. Scalapino, *Phys. Rev. Lett.* **68**, 706 (1992).
[26]  P. J. Hirschfeld, W. O. Putikka, and D. J. Scalapino, *Phys. Rev. B* **50**, 10250 (1994).
[27]  W. N. Kang, H.-J.Kim, E.-M.Choi, C. U. Jung, and Sung-IK Lee, *Science* **292**, 1521 (2001).
[28]  N.Klein et al., *J. Appl. Phys.* **67**, 6940(1990).
[29]  S. B. Nam, *Phys. Rev.* **156**, 470 (1967).
[30]  C. P. Moca, *Phys. Rev. B* **65**, 132509 (2002).
[31]  A. A. Golubov et al, *Phys. Rev. B* **66**, 054524 (2002).
[32]  K.S.Ilin, S.A.Vitusevich, B.B.Jin, A.I.Gubin, N.Klein, M.Siegel, Proceeding of M$^2$S-Rio, p.166, 2003
[33]  D. C. Mattis and J. Bardeen, *Phys. Rev.* **111**, 412 (1958).

**Figure captions:**

Fig.1: log $\delta\lambda^{eff}$ vs. $T_c/T$ representation of the penetration depth data for two MgB$_2$ samples. The solid lines are calculated within the two-gap model in the dirty limit with parameters as described in the text. Note the different scale ranges for the two samples.

Fig.2: Temperature dependence of $\sigma_1$ at 17.9GHz for MgB$_2$ samples S1210 (squares), S1211 (circles) and a Nb thin film(triangles). The lines show the calculations described in the text. The data for S1210 and S1211 are shifted by 1 and 0.5 units for clarity.

Fig.3: Calculated temperature dependence of the two energy gaps within the two band model and parameters as described in the text. The dashed line represents $\Delta(T)=kT$. The arrow points to the intersection of the dashed line and $\Delta_\pi(T)$, at which the coherence peak appears.

**Fig.1**

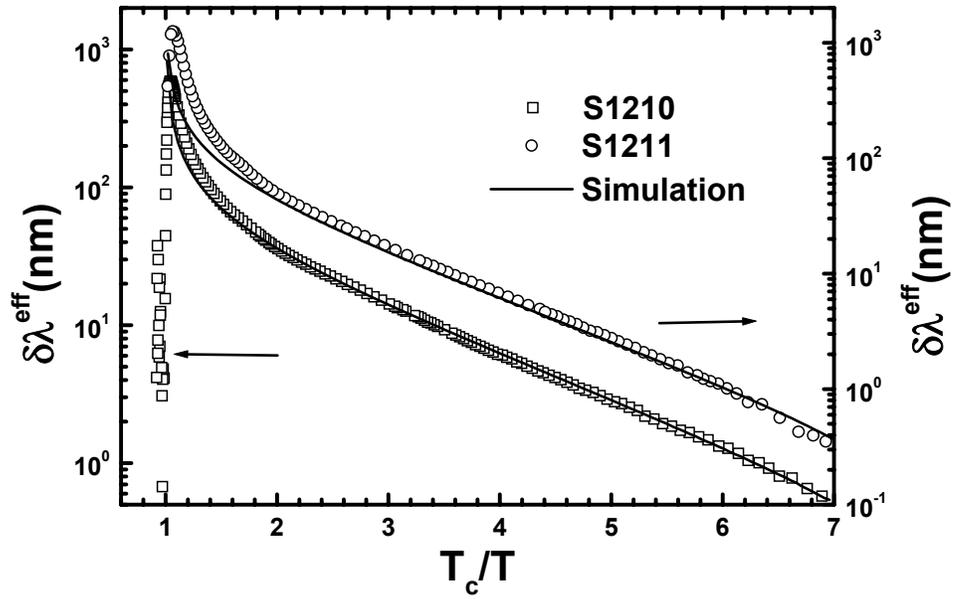

**Fig.2**

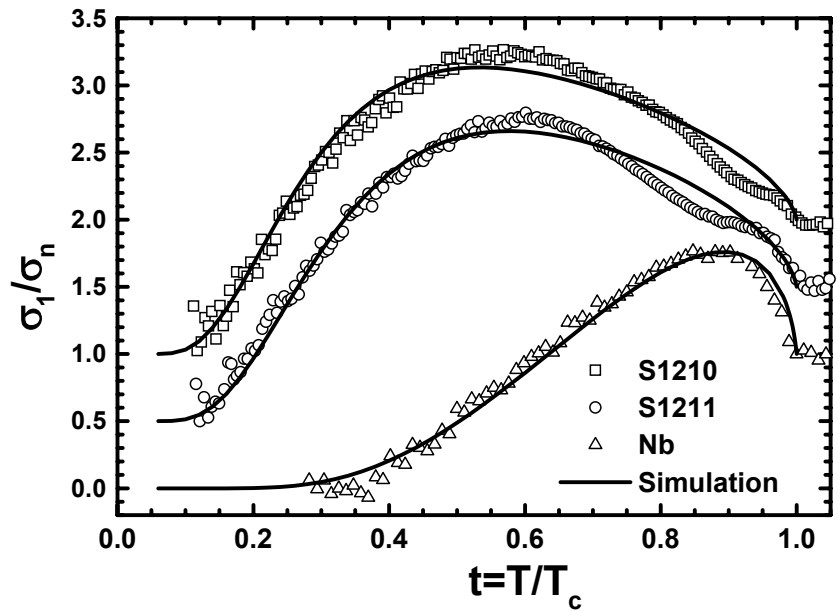

**Fig.3**

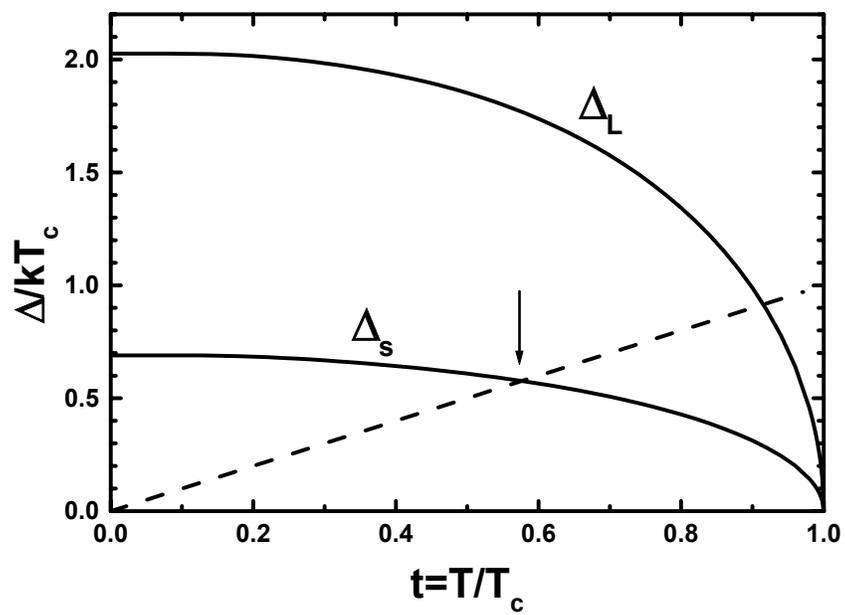